\documentclass[twocolumn, tighten]{aastex62}
\usepackage{natbib,afterpage}
\usepackage{color}
\usepackage{amssymb,times}
\usepackage[english]{babel}
\usepackage{subfigure}
\usepackage{ifpdf}
\usepackage[rightcaption]{sidecap}
\usepackage{booktabs}
\usepackage{mathtools,array}

\def\Msun{\hbox{M$_{\odot}$}}               
\def\Rstar{\hbox{R$_{\star}$}}              
\def\Mdot{\hbox{$\dot{M}$}}               

\def\deg{\hbox{$^\circ$}}       

\received{Date}
\revised{Date}
\accepted{Date}
\submitjournal{ApJ}

\shorttitle{Metal oxides and metal hydroxides in the winds of AGB stars}
\shortauthors{Decin et al.}

\begin{document}
\selectlanguage{english}
\newcommand{\red}{\textcolor[rgb]{1,0,0}}
\newcommand{\blue}{\textcolor[rgb]{0,0,1}}
\newcommand{\BlueGreen}{\textcolor[rgb]{0,1,1}}
\newcommand{\orange}{\textcolor[rgb]{1,0.5,0}}

\newlength{\fntxvi} \newlength{\fntxvii}
\newcommand{\chemical}[1]
{{\fontencoding{OMS}\fontfamily{cmsy}\selectfont
\fntxvi\the\fontdimen16\font
\fntxvii\the\fontdimen17\font
\fontdimen16\font=3pt\fontdimen17\font=3pt
$\mathrm{#1}$
\fontencoding{OMS}\fontfamily{cmys}\selectfont
\fontdimen16\font=\fntxvi \fontdimen17\font=\fntxvii}}

\title{Constraints on metal oxide and metal hydroxide abundances in the winds of AGB stars -- \\ Potential detection of FeO in R~Dor}

\correspondingauthor{Leen Decin}
\email{leen.decin@kuleuven.be}

\author{L.\ Decin}
\affiliation{Instituut voor Sterrenkunde, Katholieke Universiteit Leuven, Celestijnenlaan 200D, 3001 Leuven, Belgium}
\affiliation{University of Leeds, School of Chemistry, Leeds LS2 9JT, United Kingdom}

\author{T.\ Danilovich}
\affiliation{Instituut voor Sterrenkunde, Katholieke Universiteit Leuven, Celestijnenlaan 200D, 3001 Leuven, Belgium}

\author{D.\ Gobrecht}
\affiliation{Instituut voor Sterrenkunde, Katholieke Universiteit Leuven, Celestijnenlaan 200D, 3001 Leuven, Belgium}

\author{J.\ M.\ C.\ Plane}
\affiliation{University of Leeds, School of Chemistry, Leeds LS2 9JT, United Kingdom}

\author{A.\ M.\ S.\ Richards}
\affiliation{JBCA, Department Physics and Astronomy, University of Manchester, Manchester M13 9PL, UK}

\author{C.\ A.\ Gottlieb}
\affiliation{Harvard-Smithsonian Center for Astrophysics, Cambridge, MA 02138, and School of Engineering \& Applied Sciences, Harvard
University, Cambridge, MA 02138, USA}

\author{K.\ L.\ K.\ Lee}
\affiliation{Harvard-Smithsonian Center for Astrophysics, Cambridge, MA 02138, and School of Engineering \& Applied Sciences, Harvard
University, Cambridge, MA 02138, USA}



\begin{abstract}

Using ALMA, we observed the stellar wind of two oxygen-rich Asymptotic Giant Branch (AGB) stars, IK~Tau and R~Dor, between 335 and 362\,GHz. 
One aim was to detect metal oxides and metal hydroxides (AlO, AlOH, FeO, MgO, MgOH), some of which are thought to be direct precursors of dust nucleation and growth. 
We report on the potential first detection of FeO (v=0, $\Omega$\,=\,4, J=11--10) in R~Dor (mass-loss rate \Mdot $\sim$1$\times10^{-7}$\,\Msun/yr). 
The presence of FeO in IK~Tau (\Mdot$\sim$5$\times 10^{-6}$\,\Msun/yr) cannot be confirmed due to a blend with $^{29}$SiS, a molecule that is absent in R~Dor. 
 The detection of AlO in R~Dor and of AlOH in IK~Tau was reported earlier by \citet{Decin2017arXiv170405237D}. 
 All other metal oxides and hydroxides, as well as MgS, remain undetected. 
 We derive a column density N(FeO) of $1.1\pm0.9\times10^{15}$\,cm$^{-2}$ in R~Dor, or a fractional abundance [FeO/H]$\sim$1.5$\times$10$^{-8}$ accounting for non-LTE effects. 
 The derived fractional abundance [FeO/H] is a factor $\sim$20 larger than conventional gas-phase chemical kinetic predictions. 
This discrepancy may be partly accounted for by the role of vibrationally excited OH in oxidizing Fe, or may be evidence for other currently unrecognised chemical pathways producing FeO. 
Assuming a constant fractional abundance w.r.t.\ H$_2$, the upper limits for the other metals are 
[MgO/H$_2$] $<$5.5$\times$10$^{-10}$ (R~Dor) and $<$7$\times$10$^{-11}$ (IK~Tau),  
[MgOH/H$_2$] $<$9$\times$10$^{-9}$ (R~Dor) and $<$1$\times$10$^{-9}$ (IK~Tau), 
[CaO/H$_2$] $<$2.5$\times$10$^{-9}$ (R~Dor) and $<$1$\times$10$^{-10}$ (IK~Tau), 
[CaOH/H$_2$] $<$6.5$\times$10$^{-9}$ (R~Dor) and $<$9$\times$10$^{-10}$ (IK~Tau), and
[MgS/H$_2$] $<$4.5$\times$10$^{-10}$ (R~Dor) and $<$6$\times$10$^{-11}$ (IK~Tau). The retrieved upper limit abundances for these latter molecules are in accord with the chemical model predictions.

\end{abstract}


\section{Introduction} \label{Sec:Introduction}

The gas-phase elements Ca, Fe, Mg, Si and Ti are depleted w.r.t.\ the solar abundances in diffuse clouds. 
The formation of metal oxides and metal hydroxides and of dust species is suggested as major cause for this depletion. Indeed, a variety of metal oxides and hydroxides are prominent in a wide range of temperature and density environments. The metal oxides TiO, VO, CrO, YO, ZrO are present in the atmospheres of cool M stars \citep[see, e.g.,][]{Scalo1976A&A....48..219S, Castelaz2000AJ....120.2627C}. SiO, TiO, TiO$_2$, AlO, and AlOH are detected in the winds of oxygen-rich Asymptotic Giant Branch (AGB) stars \citep[e.g.][]{Schoier2004A&A...422..651S, Decin2010A&A...516A..69D, DeBeck2015A&A...580A..36D, Kaminski2016A&A...592A..42K, DeBeck2017A&A...598A..53D,  Kaminski2017A&A...599A..59K, Decin2018}. Other metal oxides and hydroxides such as CaO, CaOH, FeOH, MgO, and MgOH have been searched in molecular clouds and stars \citep[e.g.][]{Hocking1979A&A....75..268H, Kaminski2013ApJS..209...38K, Sanchez2015A&A...577A..52S,  Quintana-Lacaci2016A&A...592A..51Q, Velilla2017A&A...597A..25V} without success. Laboratory measurements show that FeO can be formed at high temperatures \citep[$\ge$1000\,K,][]{Cheung1982JMoSp..95..213C} and hence could be abundant in the atmospheres and inner winds of AGB stars. However, until now, FeO has only been detected in interstellar space in absorption along the line of sight toward the galactic center H{\sc ii} region Sagittarius B2 Main (Sgr B2 M) \citep{Walmsley2002ApJ...566L.109W, Furuya2003}. This detection is interpreted as due to shocks associated with star formation which might liberate some fraction of gas-phase elements from the refractory grains. FeO has remained, however, undetected in stellar atmospheres and stellar winds. Here we report the first potential detection of FeO in the stellar wind of the low mass-loss rate AGB star R~Dor (see Sect.~\ref{Sec:Observations}). No spectral features of other metal oxides and hydroxides (CaO, CaOH, MgO, MgOH) and MgS have been seen in the winds of the two oxygen-rich AGB stars R~Dor and IK~Tau surveyed with ALMA. In Sect.~\ref{Sec:Results} we show that FeO only accounts for a tiny fraction of the solar iron abundance and derive upper limit abundances for the undetected metal species.In Sect.~\ref{Sec:Discussion} we discuss the derived abundances in the framework of local thermodynamic equilibrium (LTE) and pulsation shock induced non-equilibrium chemistry models for the stellar winds.

\section{Observations} \label{Sec:Observations}

\begin{figure*}
\centering\includegraphics[angle=0,width=.8\textwidth]{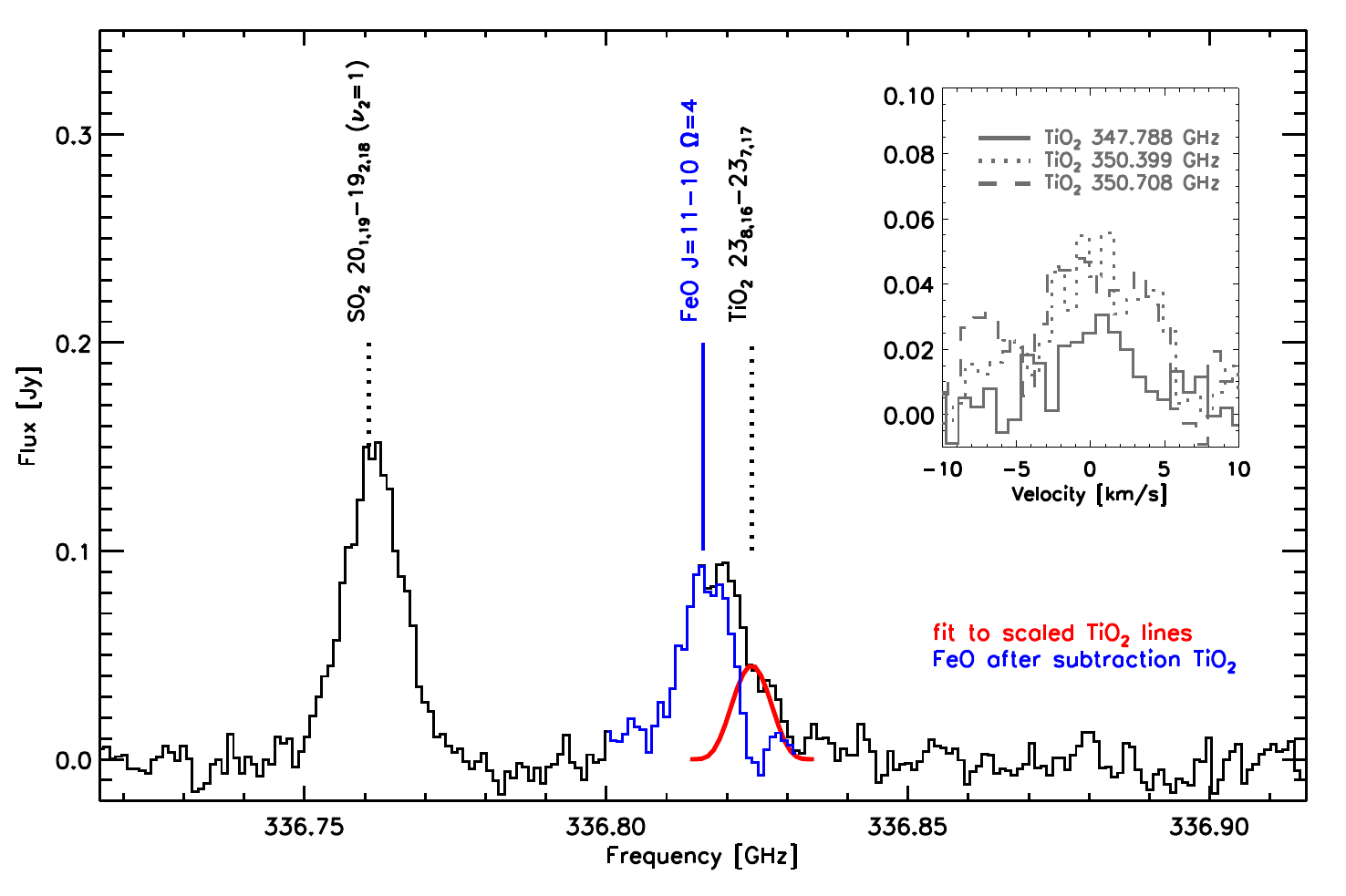}
\caption{Continuum subtracted ALMA spectrum for R~Dor around 336.8\,GHz in {\it black} extracted for a circular beam with aperture of 300\,mas. Two lines of previously identified molecules are observed: SO$_2$ ($\nu_2$=1, $20_{1,19}-19_{2,18}$), and TiO$_2$ ($23_{8,16} - 23_{7,17}$) indicated by the vertical dotted lines in {\it black}. The feature at 336.81603~GHz, tentatively identified as the  FeO ($v$\,=\,0, $\Omega$\,=\,4, $J$\,=\, 11$-$10) transition, is blended with the line of TiO$_2$ at 336.82407~GHz. The flux of the TiO$_2$ line 8~MHz higher in frequency than FeO was estimated by referring to three lines of TiO$_2$ with similar excitation energies and channel maps that were  observed in the same program (see inset in the right hand side of the panel). From these a  \textit{\red{red} synthetic} profile for the TiO$_2$ line was derived. Plotted in {\it  \color{blue} blue} is the observed profile minus the \textit{\red{synthetic}} TiO$_2$ profile attributed to FeO.}
\label{Fig:FeO}
\end{figure*}

We used ALMA to observe the high mass-loss rate  AGB star IK~Tau (\Mdot$\sim$5$\times 10^{-6}$\,\Msun/yr)  and the low mass-loss rate AGB star R~Dor (\Mdot$\sim$1$\times10^{-7}$\,\Msun/yr). Data were obtained in August-September 2015 in Band~7 (335--362\,GHz) with a spatial resolution of $\sim$150\,mas (proposal 2013.1.00166.S, PI L.\ Decin). Data reduction was done using CASA \citep{McMullin2007ASPC..376..127M} and is described in detail in \citet{Decin2018}. The spectral restoring beam parameters are in the range of 120$-$180\,mas for IK~Tau and 130$-$180\,mas for R~Dor. The channel $\sigma_{\rm{rms}}$ noise varies between spectral windows and is in the range  of 3$-$9\,mJy for IK Tau and 2.7$-$5.7\,mJy for R Dor. The velocity resolution is 1.6$-$1.7\,km/s for IK~Tau and 0.8$-$0.9\,km/s for R~Dor.

Some two hundred spectral features from 15 molecules 
were identified. 
Detected species include the gaseous precursors of dust grains such as SiO, AlO, AlOH, TiO, and TiO$_2$ \citep{Decin2017arXiv170405237D, Decin2018}. 66 lines remain unidentified, some of which may  belong to OH and H$_2$O \citep{Decin2018} or higher excitation rotational transitions not included in the current spectral line catalogues of the Jet Propulsion Laboratory \citep[JPL,][]{Pickett1998JQSRT..60..883P} and the Cologne Database for Molecular Spectroscopy \citep[CDMS,][]{Muller2001A&A...370L..49M, Muller2005JMoSt.742..215M, Endres2016JMoSp.327...95E}. The rest frequencies of the unidentified features were carefully compared to the predicted line frequencies of various metal oxides and hydroxides. Rotational transitions of CaO, CaOH, MgO, MgOH (and MgS) do not correspond to any of the unidentified lines\footnote{The rotational spectrum of FeOH in the $X ^6A^{\prime}_i$ ground state has not been measured in the laboratory. Detailed quantum mechanical calculations indicate the dipole moment of 1.368\,Debye is favourable, but FeOH is quasi-linear with a small barrier to linearity of less than 300\,cm$^{-1}$ and its spectrum may be complex \citep{Hirano2010}.}.
However, one of the unidentified spectral features in R~Dor has a central frequency around 336.815\,GHz, with a minor blend at the blue side due to a TiO$_2$ line at 366.8241\,GHz (see Fig.~\ref{Fig:FeO}). We attribute this feature to the FeO (v=0, $\Omega$\,=\,4, J=11-10) transition in the ground electronic $^5\Delta_i$ state  with rest frequency 336\,816.030$\pm$0.05\,MHz  \citep{Allen1996}. This is the only (potential) FeO line detected in our ALMA data (see also Sect.~\ref{Sec:Results}). We can not confirm if this FeO transition is present in IK~Tau due to a blend with the strong $^{29}$SiS ($v$\,=\,1, $J$\,=\,19$-$18) line at 336.815\,GHz. However, SiS (and CS) are absent\footnote{or the abundance is too low to detect even in these sensitive ALMA data} in R~Dor \citep[see][]{Decin2018}, enabling the possibility of detecting and identifying this rotational transition of FeO.

\section{Analysis and results} \label{Sec:Results}

\subsection{The FeO ($v$\,=\,0, $\Omega$\,=\,4, $J$\,=\,11$-$10) transition} \label{Sec:FeO}

The ground state of FeO is $^5\Delta_i$ in Hund's case $a$ \citep{Cheung1981JMoSp..87..289C, Cheung1982JMoSp..95..213C}. As such there are five spin-orbit components separated by intervals of 190\,cm$^{-1}$ \citep{Merer1982}, which are labelled by the quantum number $\Omega = \Lambda + \Sigma$. With $\Lambda$ and $\Sigma$ both equal to 2, vector addition gives possible values for $\Omega$ between 4 and 0, with the $^5\Delta_4$ component ($\Omega$\,=\,4) lying lowest in energy. 
The (sub)millimeter wave spectrum of FeO in the $X\,^5\Delta_i$ state (for $v$\,=\,0) has been measured up to 400\,GHz in the laboratory by \citet{Allen1996}. 

The rotational transitions for $J$\,=\,11$-$10 of all five $\Omega$ components lie in the range of the observed ALMA frequencies \citep[see Table 1 in][]{Allen1996}. Only the $J$\,=\,11$-$10 in the lowest energy $\Omega$\,=\,4 component at 336.816030\,GHz corresponds to a spectral feature in the ALMA spectrum of R~Dor (see Fig.~\ref{Fig:FeO} and the channel map in App.~\ref{App:channel_map}). 
The higher excitation $\Omega$ components remain undetected. This is unfortunate, since the detection of more than one transition of FeO would strengthen its identification. However, this outcome is not completely unexpected since transitions between the spin-orbit components are highly forbidden due to the very strong case $a$ coupling \citep{Merer1982} and hence do not support the argument that radiative transitions could cause significant transfer of population between the $\Omega$-ladders of the $X\,^5\Delta$ ground state. In this high density inner wind region, collisions might however pump the population to the higher energy $\Omega$-levels.

Other rotational transitions in the $\Omega$\,=\,4 spin-orbit component lie outside the observed ALMA frequency range\footnote{Note that in the complete ALMA archive, currently there are no data at high enough sensitivity for another low mass-loss rate AGB star that covers any of the rotational transitions in the FeO $\Omega$\,=\,4 ladder.}. The only other transition of FeO hitherto detected in interstellar space belongs to the ($v$\,=\,0, $\Omega$\,=\,4) spin-orbit component as well, and is the lowest ($J$\,=\,5$-$4) rotational transition at 153.135\,GHz towards the galactic center H{\sc ii} region Sgr B2 M \citep{Walmsley2002ApJ...566L.109W, Furuya2003}. FeO has been searched for in stellar winds since the early 1980s \citep{Merer1982} without success. If the spectral feature near 336.815\,GHz is indeed caused by FeO (and a minor blend with a TiO$_2$ line), this would be the first detection of FeO in the wind of an evolved star.

The ALMA data here offer the possibility of estimating the column density of FeO, $N$(FeO). We therefore need to subtract the contribution of the TiO$_2$ ($23_{8,16}-23_{7,17}$) transition which is slightly blended with the FeO line in the blue wing. We therefore have selected three other TiO$_2$ transitions with almost equal quantum numbers, excitation energies, line strengths, and channel maps: TiO$_2$ ($24_{2,22}-23_{3,21}$) at 347.788\,GHz, TiO$_2$ ($26_{0,26}-25_{1,25}$) at 350.399\,GHz, and TiO$_2$ ($25_{2,24}-24_{1,23}$) at 350.708\,GHz (see Fig.~\ref{Fig:FeO}). An average flux density of these three lines scaled to the same peak flux was calculated and fitted using a `soft-parabola' function (see Eq.~1 in \citet{Decin2018}; see red line in Fig.~\ref{Fig:FeO} here), that was then subtracted from the ALMA data. The resulting spectrum (for a circular aperture with beam of 300\,mas) is shown in blue in Fig.~\ref{Fig:FeO}. The peak flux of the FeO line is 0.093\,Jy and the integrated line flux is 0.79\,Jy km/s. Correcting for a local standard of rest velocity, $v_{\rm{LSR}}$, of 7\,km/s \citep{Decin2018}, the half-width of the line at zero intensity, $\Delta v$, is 7.5\,km/s. The FeO emission is essentially unresolved with the ALMA beam of $\sim$150\,mas (see the channel map in Fig.~\ref{Fig:channel_map} and discussion in App.~\ref{App:column_density}) giving us an upper limit for the detected FeO emission of radius 2.5\,\Rstar. 

\subsection{FeO column density} \label{Sec:num_dens_FeO}

We use a population diagram analysis to estimate the column density of FeO, $N$(FeO). Assuming the emission is optically thin, we derive that the column density in the upper ($J$\,=\,11) state, $ N_{\rm{u}}^{\rm{thin}}$, is 1.7$\times$10$^{13}$\,cm$^{-2}$ (see App.~\ref{App:column_density}). 
Assuming local thermodynamic equilibrium (LTE), $N$(FeO) can be calculated as
\begin{equation}
N(\rm{FeO}) = N_{\rm{u}}^{\rm{thin}}\, \frac{Q}{g_{\rm{u}}}\, \frac{1}{\exp(-E_{\rm{u}}/kT)}\,,
\label{Eq:N_FeO}
\end{equation}
with $g_{\rm{u}}$ the degeneracy of the upper rotational state, $Q$ the partition function, and $E_{\rm{u}}$ the energy level of the upper state. The energy of the $J$\,=\,11 rotational level in the $\Omega$\,=\,4 ladder ($E_u$\,=\,57.203\,cm$^{-1}$) was calculated with the spectroscopic constants in \citet{Allen1996}.
For a constant excitation temperature, $T_x$, the partition function for a linear diatomic molecule is given by \citep{Tielensbook}
\begin{equation}
Q(T_x) = \sum_i g_i \exp(-E_i/k T_x) \simeq \frac{k T_x}{h B}\,,
\label{Eq:Part}
\end{equation}
with $B$ the rotational constant of FeO in units of Hz; $B$\,=\,15493.63255\,MHz \citep{Allen1996}. 

A first estimate on the excitation temperature, $T_x$, can be obtained from calculating the (upper limit) of the cross-ladder temperature using the fact that the FeO ($v$\,=\,0, $\Omega$\,=\,3, $J$\,=\,11$-$10) at 338.844\,GHz is undetected in our survey. For an energy difference between the $\Omega$-ladders of 190\,cm$^{-1}$ \citep{Merer1982} and a $\sigma_{\rm{rms}}$ of 4\,mJy (hence 3$\sigma_{\rm{rms}}$ of 12\,mJy), we derive that the cross-ladder temperature, $T_x^{\rm{CL}}$, is $<$130\,K. If the cross-ladder populations were controlled by collisions, $T_x^{\rm{CL}}$ would be a direct measure of the kinetic temperature, $T_{\rm{kin}}$, provided cross-ladder radiative transitions are negligible, and a lower limit if not \citep{Thaddeus1984ApJ...283L..45T}. The estimated kinetic temperature in the region between 1 and 2.5\,\Rstar\ ranges between 1300--2400\,K \citep{Decin2017arXiv170405237D}. Since cross-ladder transitions are only weakly permitted (see Sect.~\ref{Sec:FeO}), the cross-ladder temperature is expected to be higher than the temperature within the ladders because populations (within a ladder) rapidly decay by emission of millimeter-wave photons  \citep{Thaddeus1984ApJ...283L..45T}. Since other rotational transitions in the $\Omega$\,=\,4 ladder are not in the frequency window of the ALMA data, we can not calculate the excitation temperature within a ladder.  \citet{Thaddeus1984ApJ...283L..45T} were able to derive the `within' and `cross'-ladder temperature for the $X^1\,A_1$ SiCC molecule --- whose $\Delta K_a=2$ electric dipole transition moments are small --- in the carbon-rich AGB star CW~Leo, being 10\,K and 140\,K respectively. Using a lower limit of 10\,K for the excitation temperature of FeO seems unreasonably low, since FeO is detected in the \textit{inner wind} of R~Dor ($r \le 2.5$\,\Rstar) in contrast to detection of SiCC in the outer wind of CW~Leo. We henceforth assume a lower limit for the excitation temperature of 80\,K. Assuming that the excitation temperature can be as high as 2000\,K, the derived column density of FeO, $N$(FeO), varies between $2 \times 10^{14}$ and $2\times 10^{15}$\,cm$^{-2}$, or $N$(FeO) is $1.1\pm0.9\times$10$^{15}$\,cm$^{-2}$.

Using the equation of mass conservation $\Mdot$\,=\,$4 \pi r^2 \rho(r) v(r)$, with $\rho(r)$ the gas density and $v(r)$ the gas velocity, one can calculate the H$_2$ density, $n$(H$_2$), assuming all hydrogen to be locked in H$_2$. The H$_2$ column density, $N$(H$_2$), is dependent on the gas velocity $v(r)$. As shown by \citet{Decin2018}, our knowledge of the gas velocity in the inner wind of R~Dor is limited. Using the velocity $\beta$-laws described by \citet{Decin2018} (their Eqs.(2)--(4)), we derive that $N$(H$_2$) is $\sim$2$\times 10^{22}$\,cm$^{-2}$ for a column with length between 1 and 2.5\,\Rstar. Hence, the ratio $N$(FeO)/$N$(H$_2$) is estimated to be $\sim$5.5$\pm 4.5\times10^{-8}$.

The FeO level populations might, however, violate the assumption of a Boltzmann distribution. There are strong transitions between the ground $X^5\Delta$ electronic state of FeO to the $^5\Pi$ and $^5\Phi$ excited electronic states near 10\,000 cm$^{-1}$ \citep{Cheung1982JMoSp..95..213C}, and owing to the large spin-orbit coupling between these three states the $X^5\Delta$ state has some excited state character that may enhance $\Delta \Omega \pm 1$ transitions between spin-components. Also rotational levels within one $\Omega$ ladder might be subject to radiative excitation effects. To check for the impact of the latter, we have calculated the frequencies, upper state energies, and Einstein A coefficients for the rotational transitions in the ($v$\,=\,0, $\Omega$\,=\,4) spin component (see App.~\ref{App:FeO_freq}). This allows us to calculate [FeO/H$_2$] by solving the statistical equilibrium equations (see Sect.~\ref{Sec:abundances}).

\subsection{Derivation of the (upper limit) abundance of FeO, MgO, MgOH, CaO, CaOH, and MgS} \label{Sec:abundances}

To derive the abundance of FeO and the upper limit abundance of the undetected species, we use the same procedure as for the determination of the AlO, AlOH and AlCl abundances in R~Dor and IK~Tau outlined in \cite{Decin2017arXiv170405237D}. In short, we modelled the ALMA data using a non-LTE radiative transfer model based on the Accelerated Lambda Iteration (ALI) method \citep{Maercker2008A&A...479..779M} that allows us to derive the global mean molecular density assuming a 1D geometry. The gas kinetic temperature and velocity have been approximated by a power law distribution. The gas density, $\rho(r)$, is calculated from the equation of mass conservation. The fractional abundance of the Al-species w.r.t.\ H$_2$ was assumed either to be constant up to a certain maximum radius, $R_{\rm{max}}$, or to decline according to a Gaussian profile centred on the star for both targets
$f(r) = f_0 \exp(-(r/R_e)^2)$, with $f_0$ the initial abundance and $R_e$ the $e$-folding radius.

Collisional excitation rates have not been published for these six molecules (as was the case for AlO, AlOH, and AlCl). Hence we have used the values from other molecules as substitutes, scaling for the difference in molecular weight. The HCN-H$_2$ system \citep{Green1974ApJ...191..653G} was used to extract the collisional rates for MgOH, CaOH (and AlOH) and the SiO-H$_2$ system \citep{Dayou2006A&A...459..297D} for FeO, CaO, MgO, MgS (and AlO, AlCl). The collisional rates are used in the form that appears in the LAMDA database\footnote{http://home.strw.leidenuniv.nl/$\sim$moldata/} \citep{Schoier2005A&A...432..369S}. Einstein-A coefficients were calculated from the quantum-mechanical line strength, $S$, as given in CDMS (MgOH, CaOH, CaO), JPL (MgS, MgO), or App.~\ref{App:FeO_freq} (FeO).

\begin{figure}[htp]
\centering\includegraphics[angle=0, width=0.4\textwidth]{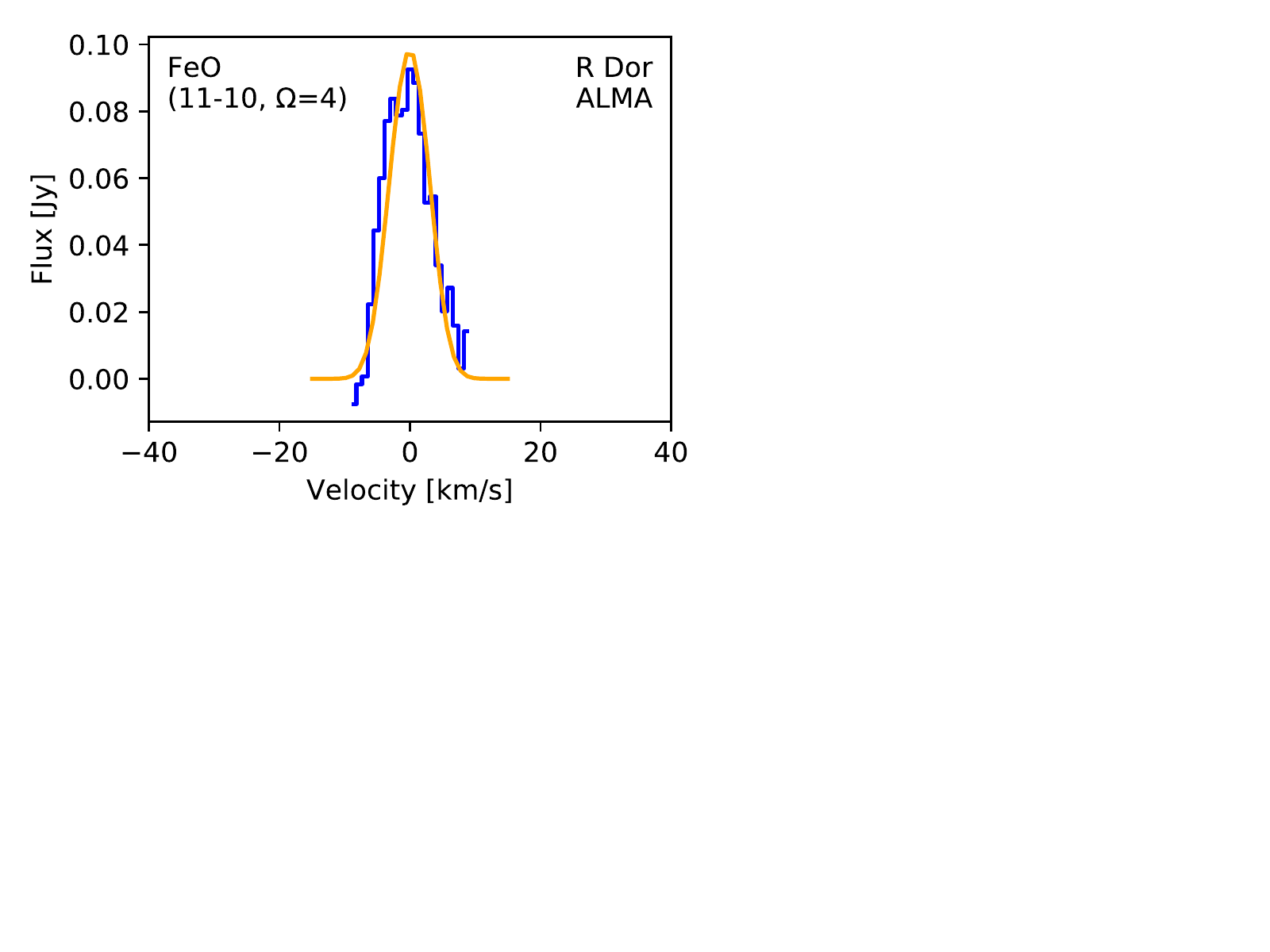}
\caption{Comparison between the \textit{\blue{blue}} extracted ALMA FeO ($v$\,=\,0, $\Omega$\,=\,4, $J$\,=\,11$-$10) spectrum (corrected for the TiO$_2$ contribution)  and the \textit{\orange{orange}} predicted line profile for [FeO/H$_2$]\,=\,$1.4\times 10^{-8}$, $R_{\rm{max}}$\,=\,6\,\Rstar, and $v_{\rm{turb}}$\,=\,3\,km/s.}
\label{Fig:FeO_fit}
\end{figure}

The ALMA R~Dor FeO channel map does not allow us to properly deconvolve the ALMA beam (see App.~\ref{App:column_density}). Assuming the ALMA emission to be essentially unresolved leads to a maximum extent of 150\,mas in diameter (or 2.5\,\Rstar\ in radius). Using the moment-0 maps would lead to a larger FeO source size with radius $\sim$6\,\Rstar. Setting $R_e$ to 2.5\,\Rstar\ \citep[by analogy with AlCl;][]{Decin2017arXiv170405237D} and $R_{\rm{max}}$ to 6\,\Rstar, the fractional abundance of FeO can be derived. However, for a turbulent velocity, $v_{\rm{turb}}$, of  1\,km/s \citep{Decin2017arXiv170405237D}, the predicted FeO line profile is smaller than observed. As described by \citet{Decin2018}, our understanding of the gas velocity in the inner wind region is limited. Pulsation-induced shocks might result in a radial velocity amplitude of a few km/s \citep{Nowotny2010A&A...514A..35N}. Allowing the turbulent velocity to be 3\,km/s permits the predicted line profile to reach the observed $\Delta v$ of 7.5\,km/s (see Fig.~\ref{Fig:FeO_fit}). Using these parameters  for the undetected species, the (upper limit) abundance of the metal species is derived (see Table~\ref{Table:abundances}). Since for IK~Tau all metal species remain undetected, the upper limit abundances were only calculated for a constant abundance profile with $v_{\rm{turb}}$\,=\,1\,km/s and $R_{\rm{max}}$\,=\,40\,\Rstar\ \citep[as determined from AlOH and AlCl,][]{Decin2017arXiv170405237D}.

\begin{table}[htbp]
\caption{Derived abundances for the metal species in R~Dor and IK~Tau.}
\label{Table:abundances}
\setlength{\tabcolsep}{2mm}
\begin{tabular}{lr@{}>{{}}lr@{}>{{}}lr@{}>{{}}l}
\hline \hline
 & \multicolumn{4}{c}{R Dor} & \multicolumn{2}{c}{IK Tau} \\
  \cmidrule(r){2-5} 
 \cmidrule(l){6-7} 
 & \multicolumn{2}{c}{constant $f$ for} & \multicolumn{2}{c}{$f_0$ for} & \multicolumn{2}{c}{constant $f$ for}  \\
\raisebox{1.5ex}[0pt]{molecule}		  & \multicolumn{2}{c}{$R_{\rm{max}}$\,=\,6\,\Rstar} & \multicolumn{2}{c}{$R_e$\,=\,2.5\,\Rstar} & \multicolumn{2}{c}{$R_{\rm{max}}$\,=\,40\,\Rstar}\\
\hline
\rule[0mm]{0mm}{5mm}[FeO/H$_2$] & 1.4 & $\times$10$^{-8}$ &       5  & $\times$10$^{-8}$ & $<6.5$ &$\times$10$^{-10}$\\
\rule[0mm]{0mm}{0mm}[MgO/H$_2$] & $<$5.5 & $\times$10$^{-10}$ & $<$4 & $\times$10$^{-9}$ & $<7$  & $\times$10$^{-11}$\\
\rule[0mm]{0mm}{0mm}[MgOH/H$_2$] & $<$9 & $\times$10$^{-9}$ & $<$4.5 & $\times$10$^{-8}$ & $<1$  & $\times$10$^{-9}$\\
\rule[0mm]{0mm}{0mm}[CaO/H$_2$] & $<$2.5 & $\times$10$^{-9}$ & $<$1.7 & $\times$10$^{-8}$ & $<1$ &  $\times$10$^{-10}$\\
\rule[0mm]{0mm}{0mm}[CaOH/H$_2$] & $<$6.5 & $\times$10$^{-9}$ & $<$3.5 & $\times$10$^{-8}$ & $<9$ &  $\times$10$^{-10}$\\
\rule[0mm]{0mm}{0mm}[MgS/H$_2$] & $<$4.5 & $\times$10$^{-10}$ & $<$2.3 & $\times$10$^{-9}$ & $<6$  & $\times$10$^{-11}$\\
\hline
\end{tabular}
\end{table}

A principal uncertainty in the abundance calculations concerns the unknown collisional rates. Changing the collisional rates by one order of magnitude only alters the retrieved abundances by 10\% or less. The only exception is a lowering by 60\% of the calculated FeO abundance if the collisional rates were a factor 10 lower.

\section{Discussion} \label{Sec:Discussion}


Both the LTE and non-LTE approach render a similar abundance [FeO/H$_2$]$\sim$3$\times$10$^{-8}$ in the inner wind of R~Dor; or [FeO/H]$\sim$1.5$\times$10$^{-8}$ assuming all hydrogen to be locked in H$_2$ for the radiative transfer calculations. In this section, we compare this derived abundance to chemical equilibrium and chemical kinetic network model predictions \citep[cf.\ earlier work done by, e.g.,][]{Gail2013pccd.book.....G, Cherchneff2006A&A...456.1001C, Gobrecht2016A&A...585A...6G}.

\subsection{Description of gas-phase chemical kinetics models} \label{Sec:description_model}

The gas-phase network used in this study contains 7 atomic (H, He, C, O, Ca, Mg, Fe) and 15 molecular (H$_2$, H$_2$O, O$_2$, OH, CO, CO$_2$, CaO, CaOH, CaH, MgO, MgOH, MgH, FeO, FeOH, and FeH) species that take part in 59 reactions. The elemental abundances are retrieved from the {\sc fruity} stellar evolution database \citep{Cristallo2015ApJS..219...40C}; dust formation is not accounted for since we focus on the region where the bulk of the dust has not yet formed.

The gas-phase reaction rate coefficients are taken from the literature where available, and extrapolated to the high temperatures of an outflow using Transition State Theory \citep[TST,][]{Atkins1998} with molecular constants (vibrational frequencies, rotational constants) calculated using quantum theory. 
The rate coefficients for reverse reactions were then calculated assuming detailed balance. The list of all chemical reactions involved is given in Table~\ref{Table:network} in App.~\ref{App:chemical_network}.

The oxides of Ca, Fe, and Mg are produced by reactions with O$_2$, CO$_2$, H$_2$O and OH releasing O, CO, H$_2$, and H, respectively, and the hydroxides of Ca, Fe, and Mg are formed by reactions with H$_2$O and OH. Moreover, the metal oxides (CaO, FeO, MgO) are linked to the hydroxides by reaction with molecular hydrogen H$_2$. 
Generally, small (reduced) networks might introduce oversimplifications compared to extensive, complete reaction networks. However, the metallic Ca-Mg-Fe chemistry is largely decoupled from the remaining gas phase chemical families (e.g.\ sulphur, nitrogen, silicon). We also compared the modelled OH (and H$_2$O) abundance with the study of \citet{Gobrecht2016A&A...585A...6G} who used an extensive chemical network with 100 species and 424 reactions (including the N, S and Si chemistry). We find similar trends and absolute values of the OH abundance in both models. The reactions R7, R8 and R11--R16 have the largest impact on the OH chemistry and determine the H$_2$O-OH balance.  In addition, the abundances of the prevalent species CO, CO$_2$, H$_2$O, and OH agree with observations.

\begin{figure}[htbp]
\centering \includegraphics[width=0.45\textwidth]{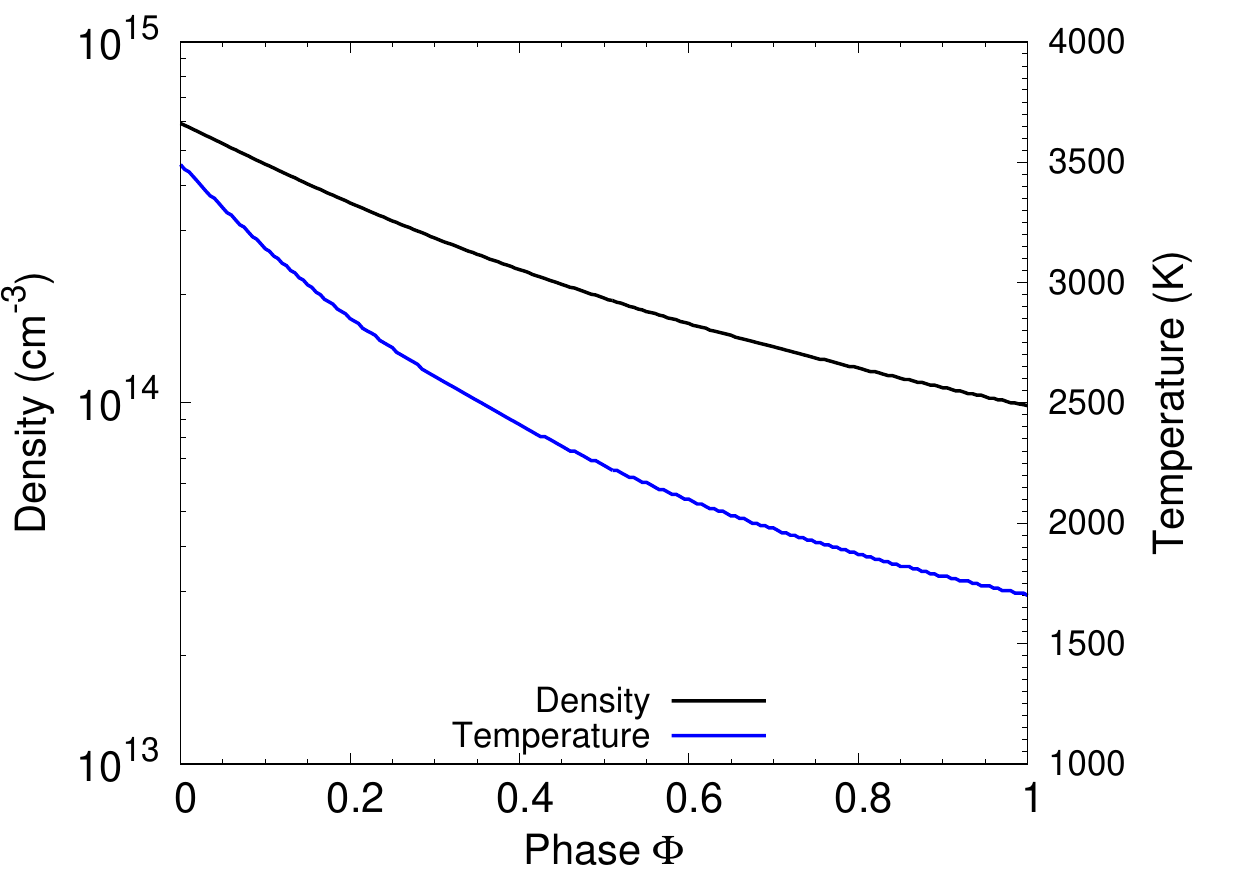}
\caption{Gas temperature and number density variation in the post-shock gas over one pulsation period at 1\,\Rstar.}
\label{Fig:DensTemp}
\end{figure}

The physical conditions experienced by the upper atmosphere of R~Dor are described by a parcel of gas which is initially at rest at the photosphere and is in thermodynamic (thermal, chemical, radiative and mechanical) equilibrium. We assume that the stellar pulsation, originating from the interior of the star, has steepened in a  shock and hits the gas parcel. As a consequence, gas in the cube is compressed, heated and accelerated outwards. The temperature and density profiles are calculated following \citet{Bertschinger1985ApJ...299..167B} for a 10\,km/s shock and a diatomic gas with pre-shock conditions of $T_0$\,=\,2400\,K and $n_0$\,=\,1$\times$10$^{14}$\,cm$^{-3}$ \citep[see, e.g., Fig.~10 in][]{Freytag2017A&A...600A.137F}. Hydrodynamic calculations \citep{Nowotny2010A&A...514A..35N} have shown that the amplitude of the velocity variation, and hence the shock velocity, is slightly larger than the terminal wind velocity \citep[being $\sim$5.5\,km/s; see discussion in ][]{Decin2018}.
At the shock front, gas temperature and density take peak values of $T$\,=\,3500\,K and $n$\,=\,6$\times$10$^{14}$\,cm$^{-3}$, respectively, and subsequently decrease with an exponential decay in the post-shock gas (see Fig.~\ref{Fig:DensTemp}). We ran the gas-phase chemistry model over a full pulsation period  \citep[being 332\,days,][]{Bedding1998MNRAS.301.1073B} and followed the change in the atomic and molecular abundance profiles over one pulsation phase (with the phase defined as the decimal part of $(t-T_0)/P$, with $T_0$ the epoch of the start of the pulsation cycle and $P$ the period); see  Fig.~\ref{Fig:FeO_abundances}. We note that the abundance variations within a pulsation period are much larger than the cycle-to-cycle variations in the periodic pulsation model.

To probe the accuracy of the chemical kinetics code and reaction network we performed runs at constant temperature ($T$\,=\,2400\,K) and density (1$\times$10$^{14}$\,cm$^{-3}$), and compared the results to equilibrium abundances with the same conditions. For the large majority of molecules, the differences are small and within a factor of 2.  Exceptions are O$_2$, FeOH, MgO and MgOH that differ by factors of up to 6--9. We conclude that our small network describes the chemical behaviour in the inner wind of R~Dor with sufficient accuracy. The relatively small differences between equilibrium and chemical-kinetic abundances may arise from the incomplete network or from an insufficient characterisation of some molecules (FeOH, MgOH) and their related reaction rate coefficients.

\subsection{Outcome of the gas-phase chemical kinetics models}

\begin{figure}[htp]
\includegraphics[angle=0,width=0.48\textwidth]{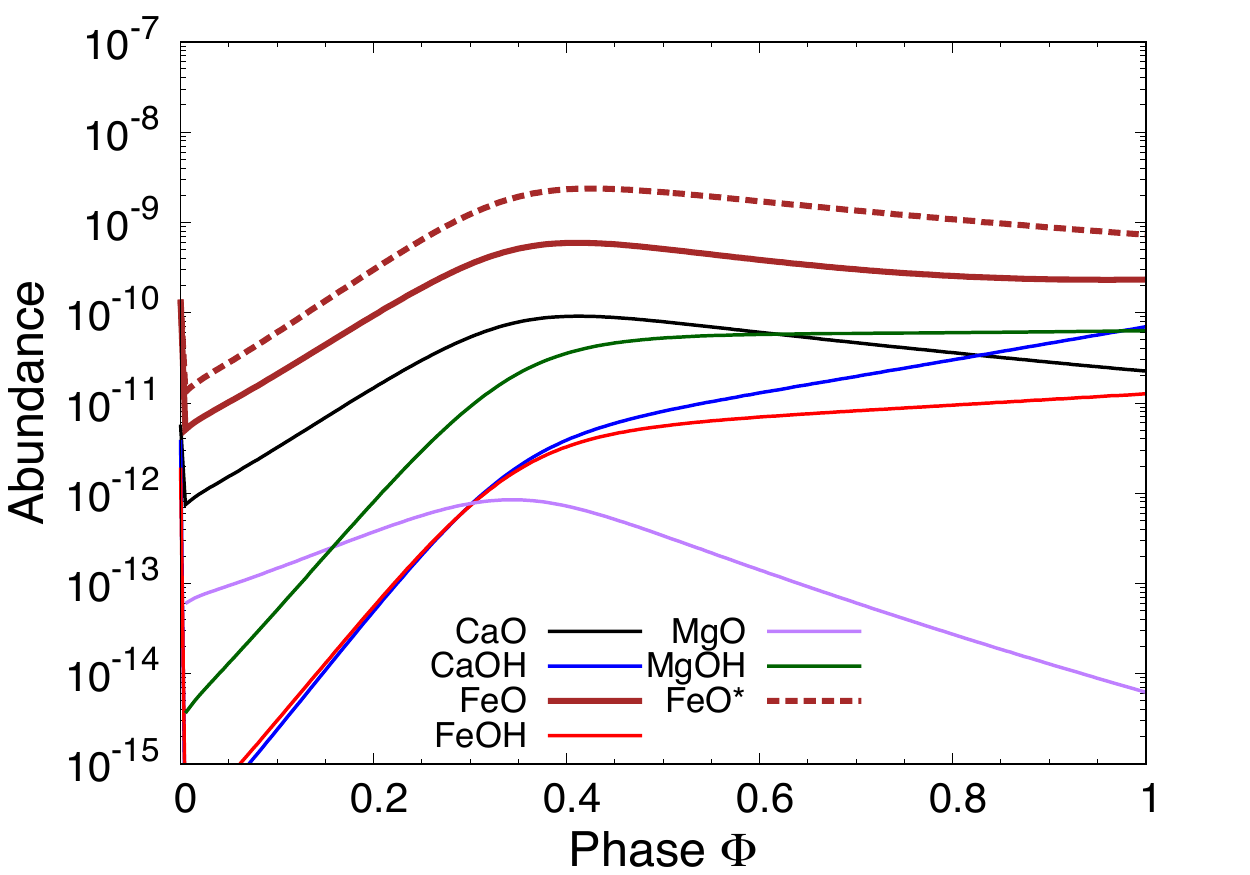}
\caption{Predicted abundances for the metal oxides and hydroxides with respect to the total gas number density as a function of pulsation phase $\phi$ at 1\,\Rstar. FeO$^\star$ corresponds to the model abundance including vibrationally excited OH molecules.}
\label{Fig:FeO_abundances}
\end{figure}

The predicted abundances for the metal oxides and metal hydroxides are shown in Fig.~\ref{Fig:FeO_abundances}. In the immediate post-shock region (pulsation phase, $\phi$, 0.0--0.2) the dissociated molecules start to reform and reach peak abundances between 1.1$\times$10$^{-12}$ ([MgO/H]) and 7.4$\times$10$^{-10}$ ([FeO/H]) in the  cooling post-shock gas ($\phi$\,=\,0.2--1.0). While the upper limit abundances for the undetected species in R~Dor (Table~\ref{Table:abundances}) are in accord with the model results, the predicted FeO abundance is a factor 20 lower than derived from the ALMA data. The main processes leading to the formation of FeO are Fe\,+\,OH (R42 in Table~\ref{Table:network}) and Fe\,+\,H$_{2}$O (R58). The first reaction dominates at early phases $\phi$\,$<$\,0.5, whereas the latter is only important at later phases $\phi$\,$>$0.5. The main FeO destruction channel is the reaction FeO\,+\,H $\rightarrow$ Fe\,+\,OH (R43).

Provided the FeO identification is correct, a number of suggestions can be put forward to explain the discrepancy between observed and predicted FeO abundance. It might be that the chemical network is not complete or that the use of detailed balance to estimate some of the rate coefficients is not correct in this environment where molecular vibrational models may not be thermally equilibrated. Another possibility is the sputtering of dust grains, although this seems unlikely since the grains close to the star should be Fe-free silicates or alumina \citep{Khouri2016A&A...591A..70K} and sputtering products such as O will actually decrease FeO (R39). Although fresh molecular O$_2$ might react with Fe, which is abundantly present ($3.1 \times 10^{-5}$ relative to H), R38 has a very large activation energy (see Table~\ref{Table:network}). We here propose an alternative scenario, following the idea of \citet{Elitzur1976ApJ...205..384E}, that the Fe\,+\,OH reaction (R42) might occur from  vibrationally excited OH, where R42 would no longer be endothermic. We account for this  possibility by reducing the activation barrier of R42 to zero. As a result, the FeO fractional abundance increases by a factor $\sim$4 (see dashed brown line in Fig.~\ref{Fig:FeO_abundances}, denoted as FeO$^\star$). 

\begin{figure}[htp]
\includegraphics[angle=0,width=0.48\textwidth]{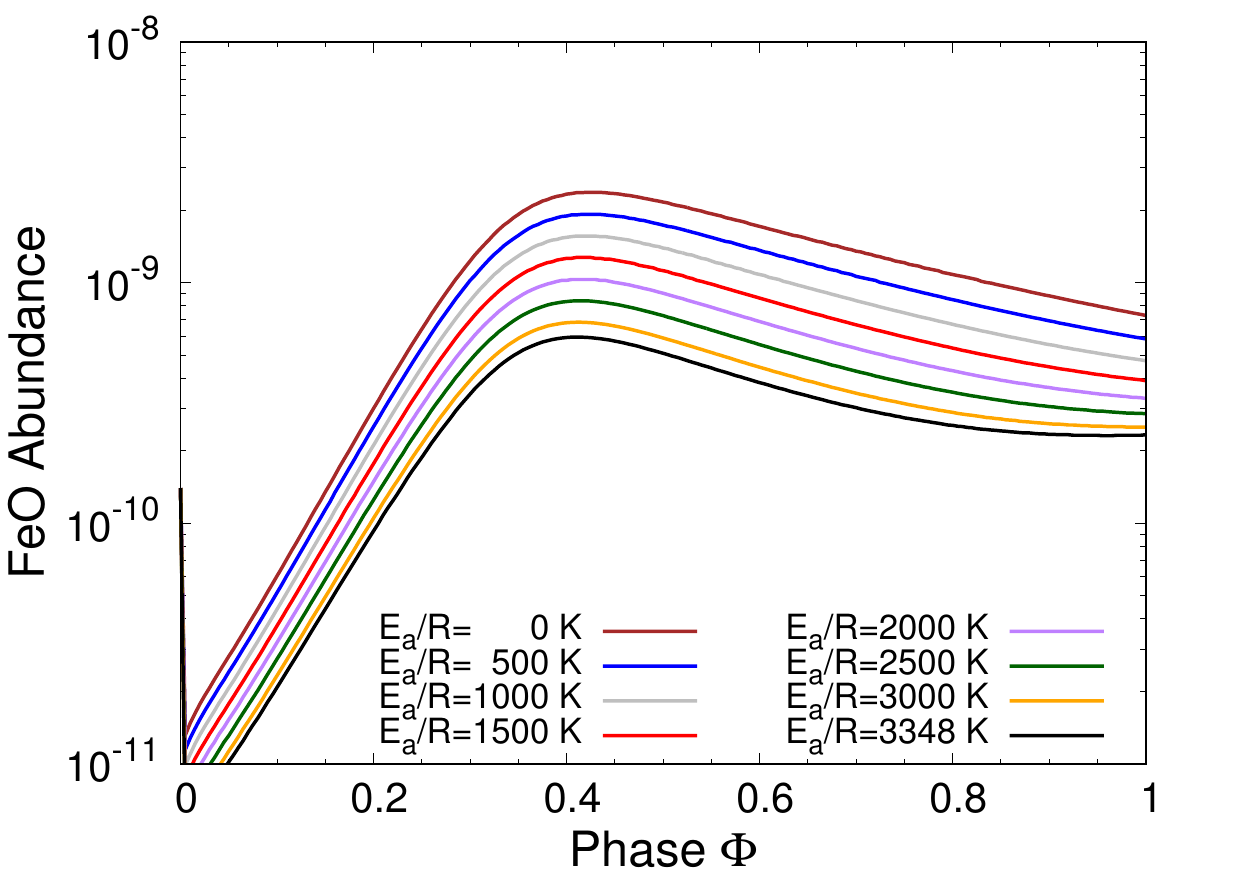}
\caption{Predicted FeO abundances with respect to the total gas number density as a function of pulsation phase $\phi$ at 1\,\Rstar\ for different values of the activation barrier, $E_a/R$,  in reaction R42.}
\label{Fig:FeO_abundances_OH}
\end{figure}

The fraction of vibrationally excited OH in the inner wind of R~Dor is, however, unknown. A first-order estimate could come from the assumption of a Boltzmann distribution of states, but this would not represent vibrational disequilibrium. The impact of the amount of vibrationally excited OH can be gauged by reducing the activation barrier, $E_a$, in reaction (R42) stepwise from 3348\,K to 0\,K, the latter situation assuming all OH is vibrationally excited hence representing an upper limit (see Fig.~\ref{Fig:FeO_abundances_OH}). As expected, the [FeO/H] maximum increases, and the maximum value reached is $2.96\times 10^{-9}$ for $E_a/R$\,=\,0\,K (see dashed brown line in Fig.~\ref{Fig:FeO_abundances} and full brown line in Fig.~\ref{Fig:FeO_abundances_OH}), which is a factor $\sim$5 lower than observed. Accounting for the uncertainties of the thermodynamics properties of the inner wind region, we conclude that this may be a viable route for the formation of gaseous FeO. However, the results also showcase that using the best available chemical kinetics, FeO is hard to make at the level tentatively observed and presents an important challenge for future chemical models.

\begin{acknowledgements}
We acknowledge A.\ Merer for the enlightening discussions concerning the FeO structure and B.\ Drouin for advice on predicting the rotational spectrum.
LD,  TD, DG,  and JMCP acknowledge support from the ERC consolidator grant 646758 AEROSOL, TD acknowledges support from the Fund of Scientific Research Flanders (FWO), JMCP from the UK Science and Technology Facilities Council (ST/P00041X/1), and CAG and KLKL from NSF grant AST-1615847. This paper uses the ALMA data ADS/JAO.ALMA2013.1.00166.S. ALMA is a partnership of ESO (representing 
its member states), NSF (USA) and NINS (Japan), together with NRC 
(Canada) and NSC and ASIAA (Taiwan), in cooperation with the Republic of 
Chile. The Joint ALMA Observatory is operated by ESO, AUI/NRAO and NAOJ.
This paper makes use of the CASA data reduction package: http://casa.nra.edu --Credit: International consortium of scientists based at the National
Radio Astronomical Observatory (NRAO), the European Southern Observatory (ESO), the National Astronomical Observatory of Japan (NAOJ), the CSIRO Australia Telescope National Facility (CSIRO/ATNF), and the Netherlands Institute for Radio Astronomy (ASTRON) under the guidance of NRAO.
\end{acknowledgements}


\bibliographystyle{aasjournal}
\bibliography{ALMA_metals}


\newpage
\begin{appendix}

\section{Channel map of the FeO (v=0, $\Omega=4$, J=11-10) transition at 336.816\,GHz}\label{App:channel_map}
\begin{figure*}[!htbp]
\includegraphics[width=\textwidth,angle=0]{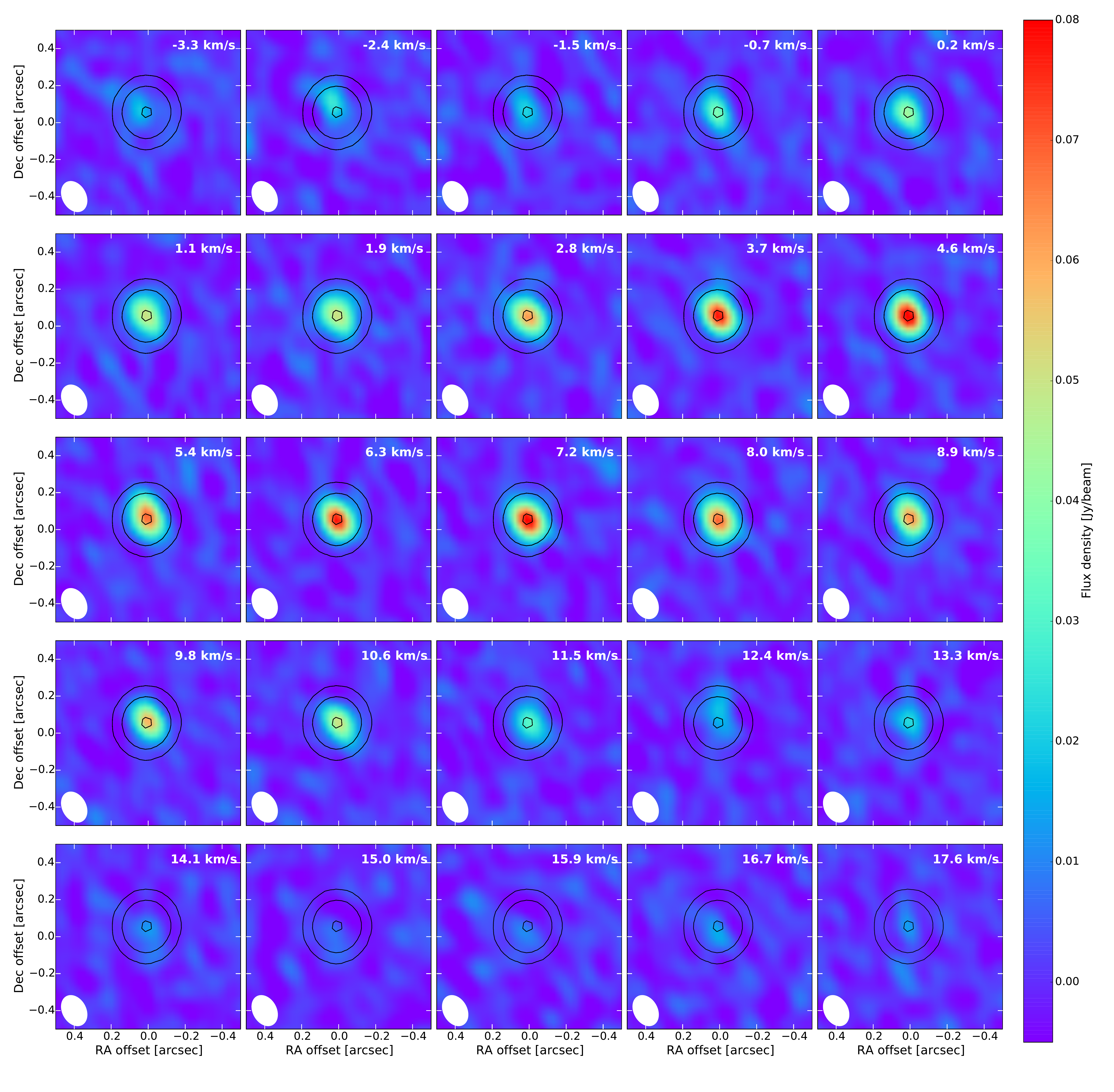}
\caption{Channel map of the FeO ($v$\,=\,0, $\Omega$\,=\,4, $J$\,=\,11--10) emission in R~Dor. The circle denotes the place of maximum dust emissivity (taking the contours at 1\%, 10\% and 90\% of the total flux). The local standard of rest velocity, $v_{\rm{LSR}}$, of R~Dor is $\sim$7\,km/s \citep{Decin2018}. The TiO$_2$ ($23_{8,16}-23_{7,17})$ transition at 266.8241\,GHz slightly blends the FeO line in the blue wing.}
\label{Fig:channel_map}
\end{figure*}

\section{Derivation of the column density of F\MakeLowercase{e}O ($\MakeLowercase{v}$\,=\,0, $\Omega$\,=\,4, J=11)-state} \label{App:column_density}

The brightness temperature $T_b$ expresses the observed intensity $I_{\nu}$ into the Planck function $B_{\nu}$
\begin{equation}
I_{\nu} = B_{\nu}(T_b)\,,
\end{equation}
which in the Rayleigh-Jeans regime simplifies to 
\begin{equation}
B_{\nu}(T_b) = \frac{2 k}{\lambda^2} \,T_b\,,
\end{equation}
with $k$ the Boltzmann constant and $\lambda$ the wavelength. The flux density $S_{\nu}^F$ of the source is defined as
\begin{equation}
S_{\nu}^F = \int_S I_{\nu}(\Omega)\, d\Omega
\end{equation}
or 
\begin{equation}
S_{\nu}^F = \frac{2 k}{\lambda^2} \int_S T_b(\Omega)\, d\Omega\,,
\label{Eq_SnuF}
\end{equation}
$\Omega$ here being the solid angle and $S$ denoting that one integrates over the source solid angle $\int_S d\Omega = \Omega_S$. For a source of uniform brightness, $T_b(\Omega)$ can be taken out of the integral and Eq.~(\ref{Eq_SnuF}) becomes
\begin{equation}
S_{\nu}^F = \frac{2 k}{\lambda^2} \,  T_b \, \Omega_S\,.
\label{Eq_SnuF_uniform}
\end{equation}

The source size $\Omega_S$ is dependent on the observed frequency $\nu$ as can be seen in Fig.~\ref{Fig:channel_map}. To derive $\Omega_S$ for each frequency, we fitted a single 2D Gaussian component to the emission in each channel from the observed frequencies between 336.7006 and 336.7455\,GHz.  The emission is, however, too compact, patchy and irregular to deconvolve the beam reliably from the apparent angular size.  This was close to the restoring beam (0$\farcs$175$\times$0$\farcs$127) size and showed large uncertainties. We hence assume that the emission is unresolved, and use the beam size ($\Omega_{\rm{beam}}$\,=\,$\Omega_S $\,=\,$0.175 \times 0.127 \times \pi/(4 \ln(2))$\,=\,0.025\,arcsec$^2$) to estimate the brightness temperature\footnote{If one used the moment-0 maps to derive the source size, one would obtain a larger angular size of $\sim$0$\farcs$360 since the emission is slightly offset from channel to channel, which makes the moment-0 image slightly enlarged. The brightness temperature derived from Eq.~(\ref{Eq_SnuF_uniform}) would be a factor 5.7 lower when using this larger source angular size.}.

Defining $W$ as
\begin{equation}
W = \int T_b\, dv\,,
\end{equation}
with $v$ the velocity (corrected for the $v_{\rm{LSR}}$)  and using Eq~(\ref{Eq_SnuF_uniform}) for an integrated flux density of the FeO ($v$\,=\,0 $\Omega$\,=\,4 $J$\,=\,11$-$10) line of 0.79\,Jy\,km/s, yields $W$\,=\,330\,K\,km/s (for a source angular size of 150\,mas in diameter or 2.5\,\Rstar\ in radius).

Assuming the emission is optically thin, one can write the column density in the upper state $N_{\rm{u}}$ as \citep{GL99}
\begin{equation}
N_{\rm{u}}^{\rm{thin}} = \frac{8 \pi k \nu^2 W}{h c^3 A_{\rm{i,j}}}\,,
\end{equation}
with $k$ the Boltzmann constant, $c$ the velocity of light, and $A_{\rm{i,j}}$ the Einstein-A coefficient for the transition. For a linear molecule, the Einstein-A coefficient for a rotational transition $J \rightarrow J-1$ is given by \citep{Tielensbook}
\begin{equation}
A_{i,j} = \frac{64 \pi^4 \nu^3 \mu_{i,j}^2}{3 h c^3}\,,
\end{equation}
with  $\mu_{i,j}$ the transition moment and $h$ the Planck constant. For a transition $J \rightarrow J-1$, the transition moment 
\begin{equation}
\mu_{i,j}^2 = \mu^2 \frac{S}{2J+1}
\end{equation}
and the quantum mechanical line strength $S$ is calculated with the standard expression for a symmetric top by replacing $K$, the angular momentum along the symmetry axis of the symmetric top, by $\Omega$ the projection of the angular momentum along the molecular axis \citep{Merer1982}
\begin{equation}
\mu^2_{J,J-1} = \mu^2 \frac{J^2 - \Omega^2}{J(2J+1)}\,,
\end{equation}
where $\mu$ is the permanent electric dipole moment. This yields
\begin{equation}
A_{J,J-1} = 1.16 \times 10^{-11} \mu^2 \nu^3 \frac{J^2 - \Omega^2}{J(2J+1)}
\label{Eq:Einstein_A}
\end{equation}
for $\nu$ in GHz and $\mu$ in Debye. Hence
\begin{equation}
N_{\rm{u}}^{\rm{thin}} = \frac{1.64 \times 10^{14}}{\nu\, \mu^2}\, \frac{J(2J+1)}{J^2-\Omega^2}\, W\,,
\end{equation}
for W in units of Jy\,km/s. Using a permanent electric dipole moment of 4.7\,Debye \citep{Steimle1989}, we obtain that $N_{\rm{u}}^{\rm{thin}}$\,=\,$1.7\times 10^{13}$\,cm$^{-2}$.

\section{Rotational transition frequencies, upper state energies, and Einstein-A coefficients in the F\MakeLowercase{e}O ($\MakeLowercase{v}$\,=\,0, $\Omega$\,=\,4)-state} \label{App:FeO_freq}

We have calculated the rotational transition frequencies and upper state energies, $E_u$, in the FeO ($v$\,=\,0, $\Omega$\,=\,4) spin component
by using the leading spectroscopic constants in \citet{Allen1996} and following \citet{Merer1982}. The Einstein-A coefficients are calculated from Eq.~(\ref{Eq:Einstein_A}); see Table~\ref{Table:FeO_table}. By comparison to the results of \citet{Allen1996}, who measured the FeO spectrum for frequencies lower than 400\,GHz, the accuracy of the calculated frequencies is about 1\,MHz. This is sufficient for our radiative transfer calculations, but we note that for spectroscopic identifications the required accuracy of the calculated frequencies in Table~\ref{Table:FeO_table} should be higher.

\begin{table}[!htbp]
\caption{Rotational transitions in the FeO ($v$\,=\,0, $\Omega$\,=\,4) ladder calculated with the spectroscopic constants of \citet{Allen1996}. Listed are the rotational frequency, upper state energy, rotational quantum numbers, and Einstein A coefficient.}
\vspace*{-2ex}
\label{Table:FeO_table}
\setlength{\tabcolsep}{2mm}
\renewcommand{\arraystretch}{0.72}
\begin{center}
\begin{tabular}{rrcr}
\hline \hline
\rule[-3mm]{0mm}{8mm}Frequency [MHz] &  \multicolumn{1}{c}{E$_{\rm{up}}$ [cm$^{-1}$]} &  \multicolumn{1}{c}{$J\rightarrow J-1$} & \multicolumn{1}{c}{$A_{J,J-1}$ [s$^{-1}$]} \\
\hline
 \rule[0mm]{0mm}{5mm}153135.0938 &    5.1080 &  5$\rightarrow$ 4 & 1.5058e-04 \\
 183757.0156 &   11.2375 &  6$\rightarrow$ 5 & 4.0768e-04 \\
 214376.1719 &   18.3883 &  7$\rightarrow$ 6 & 7.9343e-04 \\
 244992.0938 &   26.5604 &  8$\rightarrow$ 7 & 1.3299e-03 \\
 275604.3125 &   35.7535 &  9$\rightarrow$ 8 & 2.0391e-03 \\
 306212.4062 &   45.9677 & 10$\rightarrow$ 9 & 2.9429e-03 \\
 336815.8750 &   57.2027 & 11$\rightarrow$10 & 4.0635e-03 \\
 367414.2500 &   69.4583 & 12$\rightarrow$11 & 5.4226e-03 \\
 398007.1250 &   82.7344 & 13$\rightarrow$12 & 7.0422e-03 \\
 428594.0312 &   97.0307 & 14$\rightarrow$13 & 8.9441e-03 \\
 459174.5000 &  112.3471 & 15$\rightarrow$14 & 1.1150e-02 \\
 489748.0938 &  128.6834 & 16$\rightarrow$15 & 1.3682e-02 \\
 520314.3750 &  146.0392 & 17$\rightarrow$16 & 1.6561e-02 \\
 550872.8750 &  164.4143 & 18$\rightarrow$17 & 1.9810e-02 \\
 581423.1250 &  183.8085 & 19$\rightarrow$18 & 2.3449e-02 \\
 611964.7500 &  204.2215 & 20$\rightarrow$19 & 2.7501e-02 \\
 642497.2500 &  225.6529 & 21$\rightarrow$20 & 3.1987e-02 \\
 673020.2500 &  248.1024 & 22$\rightarrow$21 & 3.6927e-02 \\
 703533.2500 &  271.5698 & 23$\rightarrow$22 & 4.2345e-02 \\
 734035.8125 &  296.0546 & 24$\rightarrow$23 & 4.8260e-02 \\
 764527.5625 &  321.5565 & 25$\rightarrow$24 & 5.4694e-02 \\
 795008.0625 &  348.0751 & 26$\rightarrow$25 & 6.1668e-02 \\
 825476.8750 &  375.6100 & 27$\rightarrow$26 & 6.9204e-02 \\
 855933.6250 &  404.1609 & 28$\rightarrow$27 & 7.7322e-02 \\
 886377.7500 &  433.7273 & 29$\rightarrow$28 & 8.6043e-02 \\
 916809.0000 &  464.3087 & 30$\rightarrow$29 & 9.5387e-02 \\
 947226.8750 &  495.9048 & 31$\rightarrow$30 & 1.0538e-01 \\
 977631.0625 &  528.5151 & 32$\rightarrow$31 & 1.1603e-01 \\
1008021.0000 &  562.1390 & 33$\rightarrow$32 & 1.2737e-01 \\
1038396.4375 &  596.7762 & 34$\rightarrow$33 & 1.3942e-01 \\
1068756.8750 &  632.4261 & 35$\rightarrow$34 & 1.5219e-01 \\
1099102.0000 &  669.0882 & 36$\rightarrow$35 & 1.6571e-01 \\
1129431.5000 &  706.7620 & 37$\rightarrow$36 & 1.8000e-01 \\
1159744.7500 &  745.4469 & 38$\rightarrow$37 & 1.9507e-01 \\
1190041.6250 &  785.1424 & 39$\rightarrow$38 & 2.1095e-01 \\
1220321.6250 &  825.8480 & 40$\rightarrow$39 & 2.2766e-01 \\
1250584.3750 &  867.5630 & 41$\rightarrow$40 & 2.4521e-01 \\
1280829.5000 &  910.2868 & 42$\rightarrow$41 & 2.6363e-01 \\
1311056.7500 &  954.0190 & 43$\rightarrow$42 & 2.8294e-01 \\
1341265.6250 &  998.7588 & 44$\rightarrow$43 & 3.0315e-01 \\
\hline
\end{tabular}
\end{center}
\end{table}

\section{Reactions used in the chemical kinetics code}\label{App:chemical_network}

The gas-phase chemical network used in this study contains 7 atomic (H, He, C, O, Ca, Mg, Fe) and 15 molecular (H$_2$, H$_2$O, O$_2$, OH, CO, CO$_2$, CaO, CaOH, CaH, MgO, MgOH, MgH, FeO, FeOH, and FeH) species that take part in 59 reactions. The gas-phase reaction rate coefficients are taken from the literature where available, and extrapolated to the high temperatures of an outflow using Transition State Theory \citep[TST,][]{Atkins1998} with molecular constants (vibrational frequencies, rotational constants) calculated using quantum theory. 
The rate coefficients for reverse reactions were then calculated assuming detailed balance. 

Most of the uncertainty in the TST calculations arises from the accuracy of the energy barrier. For most of this work we have calculated the equilibrium constant and then used this for detailed balance where either the forward or backward reaction has been measured (i.e., barrier not needed). In some cases, a TST expression is fitted to a measured rate constant, with the barrier calculated from quantum theory adjusted to optimise the fit. This greatly improves the accuracy of the estimated rate constants. An example of our applied methodology and resulting accuracy can, e.g., be found in \citet{Self2003PCCP....5.1407S} for reactions R38 and R39.

\begin{figure}[htp]
\includegraphics[width=0.48\textwidth]{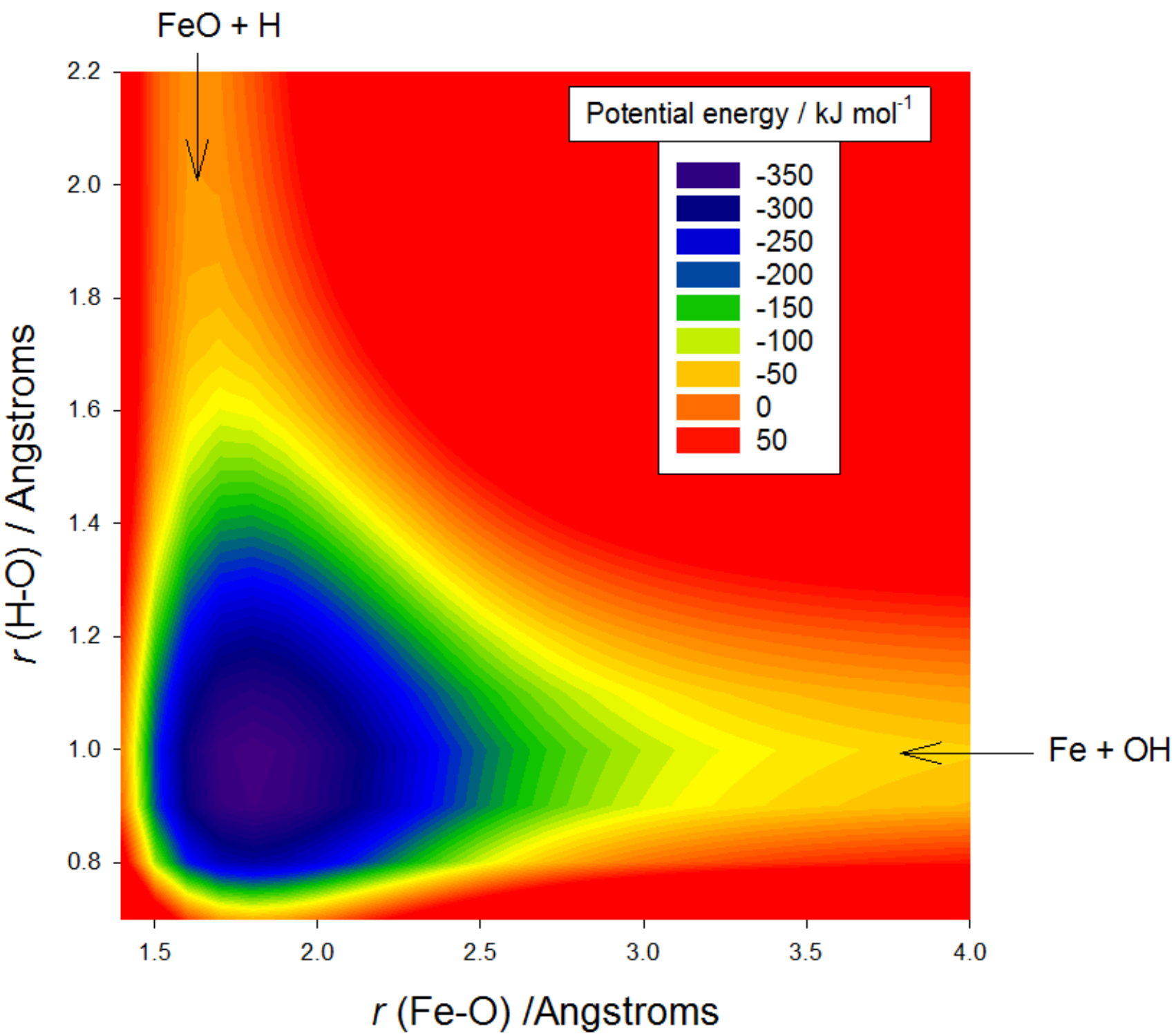}
\caption{Diagram of the potential energy surface for the Fe\,+\,OH $\rightarrow$ FeO\,+\,H reaction (R42), calculated for a fixed Fe-O-H angle of 150${\deg}$ at the b3lyp/6-311+g(2d,p) level of theory using the {\sc{Gaussian~16}} suite of programs \citep{g16}. Note the absence of barriers in the entrance channels for R42 (Fe\,+\,OH) or R43 (FeO\,+\,H). The deep well in the surface is due to the formation of FeOH.}
\label{Fig_PES}
\end{figure}

A special note concerns reactions R42 and R43. we have calculated the potential energy surface for these reactions. Fig.~\ref{Fig_PES}  illustrates the surface for a fixed Fe-O-H angle of 150$\deg$. Note there are no barriers in the entry channels of either R42 (Fe\,+\,OH) or R43 (FeO\,+\,H). We have therefore set the barrier of R43 to zero in the expression estimated by \citet{Rumminger1999}. R42 is endothermic by 25.5$\pm$5.8\,kJ\,mol$^{-1}$, using the measured FeO bond energy (0\,K) of 398.5$\pm$5.8\,kJ\,mol$^{-1}$ \citep{Metz2005JChPh.123k4313M}. The rate coefficient for R42 can then be calculated by detailed balance, yielding $k_{42}(T)$.

The list of all chemical reactions involved is given in Table~\ref{Table:network}.

\begin{table*}[!htbp]
\caption{Reaction rates for the gas-phase chemical processes used in this study\tablenotemark{a}.}
\vspace*{-3ex}
\label{Table:network}
\renewcommand{\arraystretch}{0.68}
\begin{center}
\begin{tabular}{lccccrrl}
\hline \hline
number & reactants & & products & $A$ & \multicolumn{1}{c}{$n$} & $E_a$ & Reference \& Comments\tablenotemark{b}\\
\hline
R1 & H\,+\,H\,+\,H$_2$ & $\rightarrow$ & H$_2$\,+\,H$_2$ & 8.85e$-$33 & $-$0.60 & 
    0.0 & NIST \\
R3 & H\,+\,H\,+\,H & $\rightarrow$ & H$_2$\,+\,H & 8.82e$-$33 &  0.00 &     0.0 & 
NIST \\
R5 & H\,+\,H\,+\,He & $\rightarrow$ & H$_2$\,+\,He & 4.96e$-$33 &  0.00 &     0.0
 & NIST \\
R7 & OH\,+\,OH & $\rightarrow$ & H$_2$O\,+\,O & 1.65e$-$12 &  1.10 &    50.5 & 
NIST \\
R8 & O\,+\,H$_2$O & $\rightarrow$ & OH\,+\,OH & 1.84e$-$11 &  0.95 &  8573.7 & 
NIST \\
R9 & OH\,+\,CO & $\rightarrow$ & CO$_2$\,+\,H & 3.52e$-$12 &  0.00 &  2630.2 & 
NIST \\
R10 & H\,+\,CO$_2$ & $\rightarrow$ & OH\,+\,CO & 2.51e$-$10 &  0.00 & 13229.1 & 
NIST \\
R11 & OH\,+\,H$_2$O\,+\,H & $\rightarrow$ & H$_2$O\,+\,H$_2$O & 1.19e$-$30 & $-$2.10
 &     0.0 & NIST \\
R13 & OH\,+\,H & $\rightarrow$ & H$_2$\,+\,O & 6.86e$-$14 &  2.80 &  1949.5 & 
NIST \\
R14 & O\,+\,H$_2$ & $\rightarrow$ & OH\,+\,H & 3.44e$-$13 &  2.67 &  3159.3 & 
NIST \\
R15 & H$_2$\,+\,OH & $\rightarrow$ & H$_2$O\,+\,H & 1.55e$-$12 &  1.60 &  1659.7
 & NIST \\
R16 & H\,+\,H$_2$O & $\rightarrow$ & H$_2$\,+\,OH & 6.82e$-$12 &  1.60 &  9719.8
 & NIST \\
R17 & C\,+\,O & $\rightarrow$ & CO & 1.58e$-$17 &  0.34 &  1297.4 & \citet{Dalgarno1990ApJ...349..675D} \\
R18 & C\,+\,O\,+\,M & $\rightarrow$ & CO\,+\,M & 2.00e$-$34 &  0.00 &     0.0 & 
NIST \\
R19 & CO\,+\,O\,+\,M & $\rightarrow$ & CO$_2$\,+\,M & 1.20e$-$32 &  0.00 &  2160.0
 & NIST \\
R20 & H\,+\,O\,+\,M & $\rightarrow$ & OH\,+\,M & 4.36e$-$32 & $-$1.00 &     0.0 & 
NIST \\
R21 & OH\,+\,H\,+\,M & $\rightarrow$ & H$_2$O\,+\,M & 2.59e$-$31 & $-$2.00 &     0.0
 & NIST \\
R26 & Ca\,+\,H$_2$O & $\rightarrow$ & CaO\,+\,H$_2$ & 1.70e$-$09 &  0.00 &  8749.0
 & TST estimate \\
R27 & CaO\,+\,H$_2$ & $\rightarrow$ & Ca\,+\,H$_2$O & 3.40e$-$10 &  0.00 &     0.0
 & \citet{Broadley2010PCCP...12.9094B} \\
R28 & Ca\,+\,H$_2$O & $\rightarrow$ & CaOH\,+\,H & 1.90e$-$09 &  0.00 & 19110.0 & 
TST estimate \\
R29 & CaOH\,+\,H & $\rightarrow$ & Ca\,+\,H$_2$O & 1.00e$-$10 &  0.00 &     0.0 & 
\citet{Gomez-Martin2017} \\
R30 & Ca\,+\,OH & $\rightarrow$ & CaO\,+\,H & 3.60e$-$10 &  0.00 &  4785.0 & 
TST estimate \\
R31 & CaO\,+\,H & $\rightarrow$ & Ca\,+\,OH & 1.70e$-$10 &  0.00 &  3020.0 & 
estimated as R43 \\
R32 & CaO\,+\,H$_2$ & $\rightarrow$ & CaOH\,+\,H & 2.92e$-$12 &  0.00 & $-$2050.0 & 
\citet{Cotton1971} \\
R33 & CaOH\,+\,H & $\rightarrow$ & CaO\,+\,H$_2$ & 4.48e$-$12 &  0.00 &     0.0 & 
NIST \\
R34 & Ca\,+\,CO$_2$ & $\rightarrow$ & CaO\,+\,CO & 3.20e$-$09 &  0.00 & 14907.0 & 
TST estimate \\
R35 & CaO\,+\,CO & $\rightarrow$ & Ca\,+\,CO$_2$ & 1.10e$-$11 &  0.00 &     0.0 & 
estimated as R55 \\
R36 & Ca\,+\,O$_2$ & $\rightarrow$ & CaO\,+\,O & 7.30e$-$09 &  0.00 & 11028.0 & 
TST estimate\\
R37 & CaO\,+\,O & $\rightarrow$ & Ca\,+\,O$_2$ & 1.10e$-$09 &  0.00 &   421.0 & 
\citet{Broadley2010PCCP...12.9094B} \\
R38 & Fe\,+\,O$_2$ & $\rightarrow$ & FeO\,+\,O & 2.09e$-$10 &  0.00 & 10200.0 & 
NIST \\
R39 & FeO\,+\,O & $\rightarrow$ & Fe\,+\,O$_2$ & 4.60e$-$10 &  0.00 &   350.0 & 
\citet{Self2003PCCP....5.1407S} \\
R40 & Fe\,+\,H$_2$O & $\rightarrow$ & FeO\,+\,H$_2$ & 8.80e$-$11 &  0.00 & 11925.0 & 
TST estimate \\
R41 & FeO\,+\,H$_2$ & $\rightarrow$ & Fe\,+\,H$_2$O & 2.60e$-$11 &  0.00 &  5384.0 & 
TST estimate \\
R42 & Fe\,+\,OH & $\rightarrow$ & FeO\,+\,H & 2.40e$-$10 &  0.00 &  3348.0 & 
detailed balance \\
R43 & FeO\,+\,H & $\rightarrow$ & Fe\,+\,OH & 1.70e$-$10 &  0.00 & 0.00 & 
\citet{Rumminger1999} \\
R44 & Fe\,+\,CO$_2$ & $\rightarrow$ & FeO\,+\,CO & 2.00e$-$09 &  0.00 & 16670.0 & 
TST estimate \\
R45 & FeO\,+\,CO & $\rightarrow$ & Fe\,+\,CO$_2$ & 1.20e$-$13 &  2.31 &   820.0 & 
TST estimate \\
R46 & Fe\,+\,Fe\,+\,M & $\rightarrow$ & Fe$_2$\,+\,M & 1.12e$-$31 & $-$0.52 &  7454.0 & 
NIST \\
R48 & Mg\,+\,H$_2$O & $\rightarrow$ & MgO\,+\,H$_2$ & 5.80e$-$10 &  0.00 & 29435.0 & 
TST estimate\\
R49 & MgO\,+\,H$_2$ & $\rightarrow$ & Mg\,+\,H$_2$O & 1.20e$-$10 &  0.00 &  1700.0
 & TST estimate\\
R50 & Mg\,+\,H$_2$O & $\rightarrow$ & MgOH\,+\,H & 1.10e$-$08 &  0.00 & 23173.0 & 
TST estimate \\
R51 & MgOH\,+\,H & $\rightarrow$ & Mg\,+\,H$_2$O & 1.00e$-$10 &  0.00 &     0.0 & 
estimated as R29 \\
R52 & Mg\,+\,OH & $\rightarrow$ & MgO\,+\,H & 3.40e$-$10 &  0.00 & 23651.0 & 
TST estimate \\
R53 & MgO\,+\,H & $\rightarrow$ & Mg\,+\,OH & 1.70e$-$10 &  0.00 &  3020.0 & 
estimated as R43\\
R54 & Mg\,+\,CO$_2$ & $\rightarrow$ & MgO\,+\,CO & 3.00e$-$09 &  0.00 & 33054.0 & 
TST estimate \\
R55 & MgO\,+\,CO & $\rightarrow$ & Mg\,+\,CO$_2$ & 1.10e$-$11 &  0.00 &     0.0 & 
TST estimate \\
R56 & Mg\,+\,O$_2$ & $\rightarrow$ & MgO\,+\,O & 3.00e$-$09 &  0.00 & 26584.0 & 
TST estimate  \\
R57 & MgO\,+\,O & $\rightarrow$ & Mg\,+\,O$_2$ & 6.20e$-$10 &  0.00 &     0.0 & 
\citet{Plane2012JChPh.137a4310P} \\
R58 & Fe\,+\,H$_2$O & $\rightarrow$ & FeOH\,+\,H & 1.90e$-$10 &  0.00 & 18685.0 & 
TST estimate \\
R59 & FeOH\,+\,H & $\rightarrow$ & Fe\,+\,H$_2$O & 1.00e$-$10 &  0.00 &     0.0 & 
estimated as R29 \\
\hline
\end{tabular}
\end{center}
\tablenotetext{a}{The rates are given in the Arrhenius form $k(T) = A \times \left(\frac{T}
{300}\right)^n \times \exp(-E_a/T)$, where $T$ is the gas temperature, $A$ the Arrhenius 
coefficient in cm$^3$\,s$^{-1}$ or cm$^6$\,s$^{-1}$ for a bimolecular or termolecular 
process, respectively, $n$ the temperature dependence of the rate coefficient, and $E_a$ is the activation 
energy barrier in K.}
\tablenotetext{b}{NIST: National Institute for Standards and Technology (http://
kinetics.nist.gov), TST estimate: estimation of the rate by Transition State Theory.}
\end{table*}

\end{appendix}

\end{document}